\documentclass{elsart}

\usepackage{amssymb}

\usepackage[dvips,final]{graphicx}
\usepackage{epsfig}
\usepackage{amsmath}

\begin{document}

\begin{frontmatter}



\title{LCSR analysis of exclusive two-body $B$ decay into charmonium}


\author{Bla\v zenka Meli\'c}
\ead{melic@thphys.irb.hr}

\address{Rudjer Bo\v skovi\'c Institute, Theoretical Physics Division,
HR--10002, Zagreb, Croatia
}

\begin{abstract}
We analyze $B \to K H_c$ (where $H_c = \eta_c, J/\psi, \chi_{cJ(J=0,1)}$) decays
estimating nonfactorizable contributions from the light-cone sum rules (LCSR). 
Nonfactorizable contributions are sizable for $B \to K J/\psi$ and 
particularly for $B \to K \chi_{c1}$ decay. For the $B$ decays into a 
(pseudo)scalar charmonia we find small nonfactorizable contributions which 
cannot accommodate relatively large branching ratios obtained by measurements. 
This specially concerns the puzzling $B \to K \chi_{c0}$ decay. 
\end{abstract}

\begin{keyword}

\PACS 13.25.Hw, 12.39.St, 12.38.Lg
\end{keyword}
\end{frontmatter}

\section{Introduction}
\label{Intro}
Last years there was a considerable progress in the
measurements of $B$ decays into diverse charmonium final states.
These decays governed by a color-suppressed $b \to c$ transition 
could provide a valuable information on the factorization properties of $B$-meson  
decays. 

The first observation of $B^- \to K^- \chi_{c0}$ decay a year ago by Belle
collaboration \cite{exp1}, Table \ref{tab:exp}, clearly shows the breakdown of the naive 
factorization assumption in the color-suppressed $B$ decays into charmonium. Namely, 
this decay, and the corresponding $B^- \to K^-\chi_{c2}$ decay are 
forbidden in the factorization approach 
\cite{BaBar}, due to the V-A structure of the weak vertex, i.e. 
\begin{eqnarray}
\langle \chi_{c0} | (\overline{c} c)_{V\mp A} | 0 \rangle = 
\langle \chi_{c2} | (\overline{c} c)_{V\mp A} | 0 \rangle = 0 \, . 
\label{eq:fact}
\end{eqnarray}
More surprisingly the branching ratio of $B^- \to K^- \chi_{c0} $ is 
comparable with the branching ratios of the decays $B^- \to K^- J/\psi $  and 
$B^- \to K^-\chi_{c1} $, which posses the nonvanishing factorizable amplitudes. 
Another nonfactorizable $B$ decay into $\chi_{c2}$ charmonium was observed
inclusively with a large branching fraction 
\begin{eqnarray}
{\mathcal{B}}(B \to X \chi_{c2}) = (1.80 ^{\textstyle +0.23}_{\textstyle -0.28}
\pm 0.26) \cdot 10^{-3} \, , 
\label{eq:exp}
\end{eqnarray}
which is of the order of the branching fraction of the factorizable
$B \to X \chi_{c1} $ decay. 

However, even for the $B$ decays into charmonium states which can be calculated by 
the factorization assumption 
there is a problem of theoretically too low predictions which cannot accommodate the experimental 
data. Therefore, for all $B$ decays into charmonium states one expects large nonfactorizable effects. 

\begin{table}
\label{tab:exp}
\caption{Experimental summary on branching ratios of $B$ decays into charmonium \protect{\cite{exp1}}.
 The first result for a particular decay mode is Belle 
 result, the second one is from BaBar Collaboration, and the third one (when exists) is from the 
CLEO experiment.}
\begin{tabular}{c|c} \hline
\rule[-3mm]{0mm}{8mm}
 decay mode  & $\mathcal{B} (10^{-4}$)\\
\hline\hline
\raisebox{-2.5ex}[0pt]{$B^- \to K^- \eta_c$} 
 &  $12.5 \pm 1.4 ^{+1.0}_{-1.2} \pm 3.8 $ \\ 
 &  $15.0 \pm 1.9 \pm 1.5 \pm 4.6 $ \\
 &  $6.9 ^{+2.6}_{-2.1} \pm 0.8 \pm 2.0 $ \\
\hline
\raisebox{-1.5ex}[0pt]{
$B^- \to K^- J/\psi$} 
 &  $10.1 \pm 0.3 \pm 0.8$\\
 &  $10.1 \pm 0.3 \pm 0.5$\\
\hline
$B^- \to K^- \chi_{c0}$ 
 &  $6.0 ^{+2.1}_{-1.8} \pm 1.1$\\
 &  $2.7 \pm 0.7$\\
\hline
\raisebox{-1.5ex}[0pt]{
$B^- \to K^- \chi_{c1}$} 
 & $ 6.1 \pm 0.6 \pm 0.6$ \\
 & $ 7.5 \pm 0.8 \pm 0.8$\\
\hline
\hline
\rule[-3mm]{0mm}{8mm}
$\frac{\textstyle {\mathcal{B}}(B^- \to K^- \chi_{c0})}{ \textstyle {\mathcal{B}}(B^- \to K^- J/\psi )}$ 
 & $0.60 ^{\textstyle +0.21}_{\textstyle -0.18} \pm 0.05 \pm 0.08$
\\
\hline
\end{tabular}
\end{table}

In this letter we would like to approach the calculation of the nonfactorizable 
contributions to the $B$ decay branching ratios into charmonium from the LCSR point of view.
The method was developed for the calculation of the soft nonfactorizable
corrections in $B \to \pi\pi$ \cite{khodja} and was later extended
on the $B \to K J/\psi$ decay \cite{melic}. Here we will made a short
presentation of the method, and the interested reader should have a look
in the above references for the further details.

\begin{table}
\label{tab:proper}
\caption{Charmonium states considered in the paper and their properties.}
\begin{tabular}{c||c|c|ccc} \hline
\rule[-3mm]{0mm}{8mm}
$H_c(J^{PC})$ & $j_{H_c}$ & $\langle 0| j_{H_c} | H_c \rangle $ & 
$m_{H_c}$[GeV]\protect{\cite{PDG}} & $f_{H_c}$[MeV] & $\sqrt{s_{H_c}}$[GeV]\protect{\cite{reinders}}\\
\hline\hline
$\eta_c(0^{-+})$ & $i \overline{c} \gamma_{5} c$ & $ f_{\eta_{c}} m_{\eta_{c}}$ & 3.0 &  $420 \pm 50$
\protect{\cite{hwangkim}}  & $3.8 \pm 0.2$\\
\hline 
$J/\psi(1^{--})$ & $\overline{c} \gamma_{\mu} c$ & $f_{J/\psi} m_{J/\psi} \epsilon_{\mu}$ 
& 3.1 & 405 $\pm$ 14 \protect{\cite{melic}} & $3.8 \pm 0.2$ \\
\hline
$\chi_{c0}(0^{++})$ & $\overline{c} c$ & $ f_{\chi_{c0}} m_{\chi_{c0}}$  & 3.4 &  360 \protect{\cite{novikov}}
 & $4.0 \pm 0.2$ \\
\hline
$\chi_{c1}(1^{++})$ &  $ \overline{c} \gamma_{\mu} \gamma_5 c$ & $i f_{\chi_{c1}} m_{\chi_{c1}} \epsilon_{\mu}$ 
& 3.5 & 335 \protect{\cite{novikov}}  & $4.0 \pm 0.2$ \\
\hline
\end{tabular}
\end{table}

The effective weak Hamiltonian relevant for our discussion on the $b \to c \overline{c} s $ transition 
\begin{eqnarray}
H_W = \frac{G_F}{\sqrt{2}} \Big \{ V_{cb} V_{cs}^{\ast} [C_1(\mu) \mathcal{O}_1 + C_1(\mu) \mathcal{O}_1 ] 
- V_{tb} V_{ts}^{\ast} \sum_i C_i(\mu) \mathcal{O}_i \Big \} + h.c.
\end{eqnarray}
contain the current-current operators 
\begin{eqnarray}
\mathcal{O}_1 = (\overline{c} b)_{V-A} (\overline{s} c)_{V-A} \; , \quad 
\mathcal{O}_2 = (\overline{c} c)_{V-A} (\overline{s} b)_{V-A} \, , 
\end{eqnarray}
and the QCD penguin operators $\mathcal{O}_{3-6}$, see the review \cite{review} for precise 
definitions. For the considerations in this paper, one can safely neglect contributions of the penguin operators and 
consider only contributions of the leading $\mathcal{O}_{1,2}$ operators. 
The $\mathcal{O}_1$ operator can be projected to a color-singlet state as 
\begin{eqnarray}
\mathcal{O}_1 = \frac{1}{N_c} \mathcal{O}_2 + 2  \tilde{\mathcal{O}}_2 
\end{eqnarray}
where $\tilde{\mathcal{O}}_2 = (\overline{c} T^a \gamma_{\mu} (1- \gamma_5) c)(\overline{s} T^a \gamma^{\mu}
(1- \gamma_5) b)$. 

For $H_c = \eta_c, J/\psi, \chi_{c0}, \chi_{c1}$, the decay rate can then be written as
\begin{eqnarray}
\Gamma(B \to K H_c) = \frac{G_F^2}{32 \pi} |V_{cb}|^2 |V_{cs}^{\ast}|^2 \frac{1}{m_B}\left ( 1 - 
\frac{m_{H_c}^2}{m_B^2} \right ) \, |\mathcal{M}_{H_c} |^2
\end{eqnarray}
and 
\begin{eqnarray}
\mathcal{M}_{H_c} &=& \mathcal{M}_{H_c}^{nonfact} + \mathcal{M}_{H_c}^{fact} 
\nonumber \\
&=& \left ( C_2(\mu) + \frac{C_1(\mu)}{3} \right ) 
\langle H_c K | \mathcal{O}_2 | B \rangle  + 
2 C_1(\mu) \langle H_c K | \tilde{\mathcal{O}}_2 | B \rangle  \, . 
\label{eq:M}
\end{eqnarray}
The first part of (\ref{eq:M}) can be calculated by factorizing the matrix element of the 
$\mathcal{O}_2$ operator as 
\begin{eqnarray}
\langle H_c K | \mathcal{O}_2 | B \rangle = \langle H_c | (\overline{c}c)_{V-A} | 0 \rangle 
\langle K | (\overline{s}b)_{V-A} | B \rangle \, , 
\label{eq:fact2}
\end{eqnarray}
and using the corresponding expressions for the $\langle H_c |(\overline{c}c)_{V-A}| 0 \rangle$ from 
Table 2. 
Note that due to the reason stated in introduction, Eq.(\ref{eq:fact}), there is no factorizable 
contribution to the $B \to K \chi_{c0}$ decay.  
The $B \to K$ matrix element from (\ref{eq:fact2}) is defined by the decomposition
\begin{eqnarray}
\langle K(q)|\overline{s}\gamma_{\mu} b |B(p+q) \rangle &=& \nonumber \\
& & \hspace{-4cm} 
(2 q + p)_{\mu} F_{BK}^+(p^2) + \frac{m_B^2 - m_K^2}{p^2}
p_{\mu}\, \left( - F_{BK}^+(p^2) + F_{BK}^0(p^2) \right )  \, , 
\label{eq:form}
\end{eqnarray}
while the estimation of the form factors $F_{BK}^+$ and $F_{BK}^0$ in the LCSR approach 
\cite{ball} gives the following values needed in our calculation:
\begin{eqnarray}
F_{BK}^+(m_{J/\psi}^2) = 0.60 \, ,   \quad F_{BK}^+(m_{\chi_{c1}}^2) = 0.74 \, , 
\quad F_{BK}^0(m_{\eta_{c}}^2) = 0.42  \, , 
\end{eqnarray}
with the theoretical uncertainty of $15 \%$.  

For a particular charmonium considered, the expression (\ref{eq:M}) can be then brought into the following 
form: 
\begin{eqnarray}
& & \hspace*{-0.7cm}\mathcal{M}_{J/\psi} =
\nonumber \\
& & \hspace*{-0.2cm} 2 \epsilon \cdot p_{K} \, f_{J/\psi} m_{J/\psi} F^+_{BK}(m_{J/\psi}^2) 
\left [ \left (C_2(\mu) + \frac{C_1(\mu)}{3} \right ) + 
2 C_1(\mu) \frac{\tilde{F}_{J/\psi}(\mu)}{F^+_{BK}(m_{J/\psi}^2)} \right ] \,, 
\nonumber  \\
& & \hspace*{-0.7cm}\mathcal{M}_{\chi_{c1}} = 
\nonumber \\
& & \hspace*{-0.2cm} i\, 2 \epsilon \cdot p_{K} \, f_{\chi_{c1}} m_{\chi_{c1}} F^+_{BK}(m_{\chi_{c1}}^2)
\left [ \left (C_2(\mu) + \frac{C_1(\mu)}{3} \right ) +
2 C_1(\mu) \frac{\tilde{F}_{\chi_{c1}}(\mu)}{F^+_{BK}(m_{\chi_{c1}}^2)} \right ] \, , 
\nonumber  \\
& & \hspace*{-0.7cm} \mathcal{M}_{\eta_{c}} = i\,  m_{B}^2 f_{\eta_{c}} F^0_{BK}(m_{\eta_{c}}^2)
\left [ \left (C_2(\mu) + \frac{C_1(\mu)}{3} \right ) +
2 C_1(\mu) \frac{\tilde{F}_{\eta_c}(\mu)}{F^0_{BK}(m_{\eta_c}^2)} \right ] \, , 
\nonumber  \\
& & \hspace*{-0.7cm}\mathcal{M}_{\chi_{c0}} = 2 C_1(\mu)  m_{B}^2 f_{\chi_{c0}} \tilde{F}_{\chi_{c0}}(\mu) \, . 
\end{eqnarray}
$\tilde{F}_{H_c = (\eta_c, J/\psi, \chi_{c0}, \chi_{c1})}$ 
represents the nonfactorizable part directly proportional to the contribution of 
the $\tilde{O}_2$ operator in (\ref{eq:M}). This contribution vanishes under the factorization assumption. 
Below we present the calculation of the nonfactorizable soft contributions in the LCSR approach.  

\section{Nonfactorizable contributions in $B \to K H_c$ ($H_c = \eta_c, J/\psi, \chi_{cJ(J=0,1)}$) 
decays from LCSR}

The starting point of the calculation using the LCSR method is the
correlation function, defined as
\begin{eqnarray}
\mathcal{F} = i^2 \int dx^4 \int dy^4 e^{-i p_{B} x + i (p_{H_c}-k) y}
\langle K(p_K)| j_{H_c}(y) \tilde{\mathcal{O}}_2(0) j_B(x) | 0 \rangle \, , 
\label{eq:corr}
\end{eqnarray}
where $p_K^2 = m_K^2 = 0$, $k^2=0$ and  $p_B^2 = m_B^2\, , p_{H_c}^2 = m_{H_c}^2$.
The interpolating current of a $B^-$ meson is given as $j_B = i m_b
\overline{b} \gamma_5 u$, whereas 
the choice of the interpolating
charmonium current has to be done according to the definite $J$, $P$, and
$C$ quantum numbers of a particular meson $H_c$. The considered charmonium 
currents, together with some properties of charmonium needed in the 
sum rule analysis are summarized in Table 2. 
For the consistency, 
we also include discussion on the $B \to K J/\psi$ decay which was already extensively 
presented in \cite{melic} and 
will only quote here the numerical result. 
For the rest of the $B$ decays into charmonium, the calculation closely 
follows the LCSR approach developed in \cite{khodja} and \cite{melic} and we refer to these references 
for all details. 
Including the twist-3 and twist-4 nonfactorizable contributions, we can write first the nonfactorizable 
contribution to the $B \to K \chi_{c1}$ decay analogously to the result for $B \to K J/\psi$, Eq.(65) in 
\cite{melic}:
\begin{eqnarray}
\tilde{F}_{\chi_{c1}}(\mu_b) &=& 
\frac{1}{ 4 \pi^2  f_{\chi_{c1}}^2}
\int_{4 m_c^2}^{s_0^{\chi_{c1}}} ds
\frac{(m_{\chi_{c1}}^2 + Q_0^2)^{n+1}}{(s + Q_0^2)^{n+1}}
\nonumber \\
& & \hspace*{-1.5cm} 
\frac{1}{2 m_{B}^2 f_{B}} \int_{u_0^B}^1 \frac{du}{u} e^{(m_B^2-(m_b^2 - m_{\chi_{c1}}^2(1-u))/u)/M^2}
\int_0^{1-\frac{4 m_c^2}{s}} \frac{dy}{2 \sqrt{y}} \, \frac{m_b}{m_B^2 -m_{\chi_{c1}}^2}
\nonumber \\
& & \hspace*{-1.5cm} \Bigg \{ 
\frac{f_{3K}}{2} \Bigg [
 \int_0^u
 \frac{dv}{v^2} \phi_{3K}(1 -u, u-v, v)
 \left ( \frac{m_b^2 - m_{\chi_{c1}}^2}{u} ( 2 v - X ) 
     + s - \frac{4 m_c^2}{1-y} \right )  \nonumber \\
 & & \hspace*{-0.5cm}  - \left ( s -  \frac{4 m_c^2}{1-y} \right ) 
\left ( \frac{1}{v^2} \phi_{3K}(1 -u, u-v, v) \right )_{v=0}
  X \Bigg ]
\nonumber \\
& & \hspace*{-1cm} +
m_b f_K \int_0^u
\frac{dv}{v} \tilde{\phi}_{\perp}(1 -u, u-v, v)
\left [ 3 - \frac{2}{v} X \right ]
 \Bigg \} \, ,
\label{eq:FtildeCHI1}
\end{eqnarray}
where $s_0^{H_c}$ and $s_0^B$ are the effective threshold parameters of the perturbative 
continuum in the $H_c$ and $B$ channel, respectively, and $u_0^B = (m_b^2 - m_{H_c}^2)/(s_0^B - m_{H_c}^2)$, 
here specified for $H_c = \chi_{c1}$ charmonium.  
$M$, the Borel parameter in the $B$ channel, and the parameter $n =1,2,...$ in the charmonium 
channel have to be chosen in such a way that a reliable perturbative 
calculation is possible, but on the other hand that excited and continuum states in a 
given channel are suppressed. 
The function $X$ appearing above is $X = x(s,y,m_B^2) = (s - 4 m_c^2/(1-y) )/(s - m_B^2)$ and the 
expansion up to $O(X^2)$ is performed. 
The twist-3, $\phi_{3K}$,  and the twist-4, $\tilde{\phi}_{\perp}$, three-particle kaon distribution amplitudes 
are defined as 
usual \cite{BF}. The scale at which $\tilde{F}_{H_c}$'s are calculated is $\mu_b \sim m_b/2 \sim 2.4$ GeV. 

As for the $B$ decays into the scalar and pseudoscalar charmonium the calculation yields
\begin{eqnarray}
\tilde{F}_{\eta_{c}}(\mu_b) &=& \frac{1}{f_{\eta_c} m_B^2} 
\frac{1}{ 4 \pi^2  f_{\eta_{c}}}
\int_{4 m_c^2}^{s_0^{\eta_{c}}} ds
\frac{(m_{\eta_{c}}^2 + Q_0^2)^{n+1}}{(s + Q_0^2)^{n+1}}
\nonumber \\
& & \hspace*{-1.5cm} 
\frac{1}{2 m_{B}^2 f_{B}} \int_{u_0^B}^1 \frac{du}{u} e^{(m_B^2-(m_b^2 - m_{\eta_{c}}^2(1-u))/u)/M^2}
\int_0^{1-\frac{4 m_c^2}{s}} \frac{dy}{(1-y) \sqrt{y}} \, \frac{m_b m_c}{m_{\eta_{c}}}
\nonumber \\
& & \hspace*{-1.5cm}
\Bigg \{ 
f_{3K} \Bigg [
 \int_0^u
 \frac{dv}{v} \phi_{3K}(1 -u, u-v, v)
 \left ( - \frac{m_b^2 - m_{\eta_{c}}^2}{u} \right )  \Bigg ]
\nonumber \\
& & \hspace*{-1cm} +
3 m_b f_K \int_0^u
\frac{dv}{v} \tilde{\phi}_{\perp}(1 -u, u-v, v)
 \Bigg \} \, ,
\label{eq:FtildeETA}
\end{eqnarray}
and 
\begin{eqnarray}
\tilde{F}_{\chi_{c0}}(\mu_b) &=& \frac{1}{f_{\chi_{c0}} m_B^2}
\frac{1}{ 4 \pi^2  f_{\chi_{c0}}}
\int_{4 m_c^2}^{s_0^{\chi_{c0}}} ds
\frac{(m_{\chi_{c0}}^2 + Q_0^2)^{n+1}}{(s + Q_0^2)^{n+1}}
\nonumber \\
& & \hspace*{-1.5cm} 
\frac{1}{2 m_{B}^2 f_{B}} \int_{u_0^B}^1 \frac{du}{u} e^{(m_B^2-(m_b^2 - m_{\chi_{c0}}^2(1-u))/u)/M^2}
\int_0^{1-\frac{4 m_c^2}{s}} \frac{dy}{(1-y) \sqrt{y}} \, \frac{m_b m_c}{m_{\chi_{c0}}}
\nonumber \\
& & \hspace*{-1.5cm} \times
\Bigg \{
f_{3K} \Bigg [
 \int_0^u
 \frac{dv}{v} \phi_{3K}(1 -u, u-v, v)
 \left (\frac{m_b^2 - m_{\chi_{c0}}^2}{u} \right )
 \left ( y + (1-y) \frac{X}{v} \right )  \Bigg ]
\nonumber \\
& & \hspace*{-1cm} +
3 m_b f_K \int_0^u
\frac{dv}{v} \tilde{\phi}_{\perp}(1 -u, u-v, v)
 \left ( y + (1-y) \frac{X}{v} \right ) 
 \Bigg \} \, .
\label{eq:FtildeCHI0}
\end{eqnarray}
\section{Numerical predictions and discussions}

Let us first specify the numerical values of the needed parameters. 
For parameters in the $B$ channel we use $m_B =  5.28$ GeV and the values taken from \cite{BKR}:
$f_B = 180 \pm 30 $ MeV, $m_b = 4.7 \pm 0.1$ GeV, and $s^{B}_0 = 35 \pm 2\, {\rm GeV}^2$. 
For the charmonium states 
we use the parameters from Table 2 
and $m_c = 1.25 \pm 0.10$. 
The $K$ meson decay constant is taken
as $f_K = 0.16$ GeV.  For parameters which enter the coefficients of the twist-3 and
twist-4 kaon wave functions we suppose that $f_{3 \pi} \simeq f_{3 K}$ and
$\delta^2_K \simeq \delta_{\pi}^2$,
and take $f_{3 K} = 0.0026$ GeV, $\delta^2(\mu_b) = 0.17$ GeV \cite{BF}. 
The stability region for the Borel parameter is found in the interval $M^2 = 10 \pm 2\, {\rm GeV}^2$,
known also from other LCSR calculations of $B$ meson properties. Concerning the sum rules in the 
charmonium channels, the calculation is rather stable on the change of $n$ in 
the interval $n = 4 - 7$.  $Q_0^2$ is parameterized by $Q_0^2 = 4 m_c^2 \xi$, 
where in order that the lowest resonances dominate $\xi$ takes values from 0.5 to 1 for the $B$ decays 
into $s$-wave charmonia, while 
$\xi$ is between 1 and 2.5 in the decays into $p$-wave charmonia \cite{reinders}. 

\begin{table}
\label{tab:th}
\caption{Theoretical results for the $B \to K H_c$ decays calculated in this paper. 
$\tilde{F}_{H_c}(\mu_b)$ is the  nonfactorizable contribution, $\mathcal{M}_{H_c}^{nonfact}
/\mathcal{M}_{H_c}^{fact}$, Eq.(\ref{eq:M}), 
is the ratio of the nonfactorizable and factorizable amplitudes for a particular mode $B \to K H_c$, whereas 
$\mathcal{B}$ is the branching
ratio, all calculated at $\mu = \mu_b$. 
Large scale-dependent uncertainties pertinent to the factorizable amplitude are not included. 
}
\begin{tabular}{c|c|c|c} \hline
\rule[-3mm]{0mm}{8mm}
 decay mode  & $\tilde{F}_{H_c}(\mu_b)$ &  $\mathcal{M}_{H_c}^{nonfact}/\mathcal{M}_{H_c}^{fact}$ 
& $\mathcal{B} (10^{-4})$\\
\hline\hline
$B^- \to K^- \eta_c$
& 0.0015-0.0019 & 0.08-0.10 & $2.0 \pm 0.1 $  \\
\hline
$B^- \to K^- J\psi$
& 0.011-0.018 \protect{\cite{melic}}&  0.40-0.70 \protect{\cite{melic}} & $ 3.3 \pm 0.6$\\
\hline
$B^- \to K^- \chi_{c0}$
& -(0.0007-0.0008) &   -  &  $ 0.0017 \pm 0.0002 $ \\
\hline
$B^- \to K^- \chi_{c1}$
& 0.044 - 0.052 & 1.30-1.50 & $ 5.1 \pm 0.5$\\
\hline
\end{tabular}
\end{table}
                                                                                                                           
The results of our calculation are summarized in Table \ref{tab:th}. 
Comparing the numbers from Table \ref{tab:exp} and Table \ref{tab:th}, it is important to note that in general 
the calculated branching ratios are still too low to accommodate the data, except maybe for 
the $B \to K \chi_{c1}$ decay. The nonfactorizable correction to $B \to K J/\psi$ and 
particularly to $B \to K \chi_{c1}$ is large, for the $B \to K \chi_{c1}$ decay this 
correction is even larger than the factorizable contribution, Table \ref{tab:th}. On the other hand, in 
$B$ decays into (pseudo)scalar charmonia the nonfactorizable contributions are small. The 
reason is the cancellation of the twist-3 and twist-4 contributions in these decays. 
However, even without this cancellation, the nonfactorizable effects produced by the exchange 
of a soft gluon between a kaon and $\chi_{c0}$ as calculated here, could not be able to 
account for such a large branching ratio of $B \to K \chi_{c0}$ as measured experimentally. These 
would demand contributions which have to be at least an order of magnitude larger than those typically 
occurring in a $B$ decay into charmonium state as estimated by the LCSR method. 
Therefore it is very unlikely that the mechanism of the nonfactorizable soft gluon exchange is the 
reason for a relatively large branching ratio of the $B \to K \chi_{c0}$  decay. 
A possible speculative explanation for large discrepancies between the theoretical 
results for the B decays into (pseudo)scalar
charmonia and the experimental data 
one could find in the nonperturbative effects 
of the instanton type. They could appear due to the light quark admixtures in the 
$\eta_c$ and $\chi_{c0}$ mesons, 
yet it would be hard to account for such contributions reliably. 

The  $B \to K \chi_{c0}$  decay was also analyzed within QCD factorization method \cite{BBNS} and 
it was observed \cite{songchao} that there is a problem of logarithmic 
divergences of the decay amplitude already at the leading-twist order.  
In another approach, by studying the mechanism of the rescattering of charmed
intermediate states in $B \to K \chi_{c0}$, the authors of Ref.\cite{col} show that 
the rescattering effects 
could provide the large part of the $B \to K \chi_{c0}$ amplitude.

To summarize, we have studied the soft nonfactorizable contributions to $B \to K H_c 
(H_c = \eta_c, J/\psi, \chi_{c0}, \chi_{c1})$ decays by using the LCSR approach. 
Inspite of the expected large 
contributions which could explain large discrepancy between the factorizable predictions and the 
experimental data, we were not able to confirm these expectations, except for the $B \to K \chi_{c1}$ 
decay, for which large nonfactorizable corrections are found. The other predicted $B$ decays into charmonium 
receive nonfactorizable soft contributions too small to accommodate the data. Unfortunately, 
this is particularly true for the puzzling $B \to K \chi_{c0}$ decay that factorized amplitude vanishes and 
the LCSR mechanism considered in this paper cannot explain its relatively large branching ratio.

{\it Note added:}
\\
While this work has been prepared, Ref. \cite{wang} appeared, where the authors discuss 
nonfactorizable soft contributions in the $B \to K \eta_c$ and $B \to K \chi_{c0}$ decays 
within the LCSR approach. 
There are some differences observed in the approach as well as in the results between \cite{wang} 
and our paper. As for 
the $B \to K \eta_c$ decay, the authors of \cite{wang} choose the pseudovector current for 
$\eta_c$. In that case, the result can be easily derived from Eq.(\ref{eq:FtildeCHI1}) 
according to the approach taken from \cite{melic}. The differences show in the twist-4 part. 
Apart from this, by taking the  pseudovector current to describe $\eta_c$ one has also to include 
the mixing with the $\chi_{c1}(1^{++})$ state explicitly \cite{AK} 
which was not considered in \cite{wang}.  
More importantly, for the problematic $B \to K \chi_{c0}$ decay, apart from the superfluous
factor $x$ in Eq.(49) of \cite{wang}, we agree analytically in the twist-3 part. 
In the second version of \cite{wang}, the authors have corrected numerical errors and have 
included the twist-4 contribution to their result for the nonfactorizable $B \to K \chi_{c0}$ amplitude. 
Although we now agree that the soft nonfactorizable contribution in the $B \to K \chi_{c0}$ decay is 
small and cannot accommodate the experimental data, we still disagree in the twist-4 part of 
the nonfactorizable contributions, which renders the numerical results from this paper 
somewhat different from those obtained in \cite{wang}. 

\ack
I am grateful to D. Be\v cirevi\'c, F. De Fazio, A. Khodjamirian, A.A. Petrov, K. Passek-Kumeri\v cki for stimulating 
discussions and to A. Khodjamirian also for carefully reading the manuscript. 
This work is supported by the Ministry of Science, Education and Sport of the Republic of Croatia 
under the contract 0098002 and by the Alexander von Humboldt Foundation.


\begin{thebibliography}{00}




\bibitem{exp1}
F.~Fang, ``B to charmonium: Mini-summary,'' in Proc. of the Flavor Physics and CP Violation (FPCP), Philadelphia, 
Pennsylvania, 16-18 May 2002, Ed. R.G.C. Oldeman, eConf C020516, 
[arXiv:hep-ex/0207004]. 


\bibitem{BaBar}
P.~F.~Harrison and H.~R.~Quinn  [BABAR Collaboration],
``The BaBar physics book: Physics at an asymmetric B factory,''
SLAC-R-0504. 


\bibitem{khodja}
A.~Khodjamirian,
Nucl.\ Phys.\ B {\bf 605} (2001) 558
[arXiv:hep-ph/0012271].
                                                                                                                           
\bibitem{melic}
B.~Meli\'c, 
Phys.\ Rev.\ D {\bf 68} (2003) 034004 
[arXiv:hep-ph/0303250].

\bibitem{review}
G.~Buchalla, A.~J.~Buras and M.~E.~Lautenbacher,
Rev.\ Mod.\ Phys.\  {\bf 68} (1996) 1125
[arXiv:hep-ph/9512380].



\bibitem{PDG}
K.~Hagiwara {\it et al.}  [Particle Data Group Collaboration],
Phys.\ Rev.\ D {\bf 66} (2002) 010001.
                                                                                                                           
\bibitem{reinders}
L.~J.~Reinders, H.~Rubinstein and S.~Yazaki,
Phys.\ Rept.\  {\bf 127} (1985) 1.
                                                                                                                           
\bibitem{hwangkim}
D.~S.~Hwang and G.~H.~Kim,
Z.\ Phys.\ C {\bf 76} (1997) 107
[arXiv:hep-ph/9703364].

\bibitem{novikov}
V.~A.~Novikov, L.~B.~Okun, M.~A.~Shifman, A.~I.~Vainshtein, M.~B.~Voloshin and V.~I.~Zakharov,
Phys.\ Rept.\  {\bf 41} (1978) 1.

\bibitem{ball}
P.~Ball,
JHEP {\bf 9809} (1998) 005
[arXiv:hep-ph/9802394]; 
A.~Khodjamirian, R.~Ruckl, S.~Weinzierl, C.~W.~Winhart and O.~I.~Yakovlev,
Phys.\ Rev.\ D {\bf 62} (2000) 114002
[arXiv:hep-ph/0001297].

\bibitem{BF}
V.~M.~Braun and I.~E.~Filyanov,
Z.\ Phys.\ C {\bf 48} (1990) 239. 


\bibitem{BKR}
V.~M.~Belyaev, A.~Khodjamirian and R.~Ruckl,
Z.\ Phys.\ C {\bf 60} (1993) 349
[arXiv:hep-ph/9305348].


\bibitem{BBNS} 
M.~Beneke, G.~Buchalla, M.~Neubert and C.~T.~Sachrajda,
Nucl.\ Phys.\ B {\bf 591} (2000) 313
[arXiv:hep-ph/0006124].

\bibitem{songchao}
Z.~Z.~Song and K.~T.~Chao,
Phys.\ Lett.\ B {\bf 568} (2003) 127
[arXiv:hep-ph/0206253].

\bibitem{col}
P.~Colangelo, F.~De Fazio and T.~N.~Pham,
Phys.\ Lett.\ B {\bf 542} (2002) 71
[arXiv:hep-ph/0207061].


\bibitem{wang}
Z.~G.~Wang, L.~Li and T.~Huang,
``Nonfactorizable soft contributions in the $B \to \eta_c K, \chi_{c0} K$ decays
with the light-cone sum rules approach,''
arXiv:hep-ph/0311296.


\bibitem{AK}
A. Khodjamirian, private communication. 







\end{thebibliography}
\end{document}